\documentclass[
reprint,
 amsmath,amssymb,
 aps,
 prl,
]{revtex4-2}

\usepackage{float}
\usepackage{mathtools}
\usepackage{graphicx}% Include figure files
\usepackage{dcolumn}% Align table columns on decimal point
\usepackage{bm}% bold math
\usepackage{upgreek}
\usepackage{lipsum, babel, xcolor}
\usepackage{amsmath, amssymb}
\usepackage{comment}
\usepackage{natbib}
\usepackage{physics}
\usepackage{hyperref}
\hypersetup{colorlinks=true,citecolor=blue}

\begin{document}

\preprint{APS/123-QED}

\title{Periodically poled thin-film lithium niobate ring Mach Zehnder coupling interferometer for efficient quantum frequency conversion}

\author{Mrinmoy Kundu$\mathrm{^{1}}$, Bejoy Sikder$\mathrm{^{1,2}}$, Mark Earnshaw$\mathrm{^{3}}$, A.  Sayem$\mathrm{^{3}}$}
 %\altaffiliation[Also at ]{Physics Department, XYZ University.}%Lines break automatically or can be forced with \\
%author{Second Author}%
%\email{ayed.sayem@nokia-bell-labs.com}
\affiliation{%
 BUET, Dhaka, Bangladesh $\mathrm{^{1}}$ \\
 Massachusetts Institute of Technology (MIT), Cambridge, MA, USA $\mathrm{^{2}}$ \\
 Nokia Bell Labs, NJ, USA $\mathrm{^{3}}$
% This line break forced with \textbackslash\textbackslash
}%

\date{\today}% It is always \today, today,
             %  but any date may be explicitly specified

\begin{abstract}
Quantum frequency conversion is unavoidable for a true quantum communication network as most quantum memories work in the visible spectrum. Here, we propose a unique design of a quantum frequency converter based on a ring-Mach Zehnder interferometer coupled with a periodically poled thin-film lithium niobate waveguide. The proposed device can be-directionally convert quantum signals i.e. single photons from quantum memory such as SiV-center in diamond to the telecom wavelength offering conversion efficiency as high as $\mathrm{90\%}$ at mW pump power with noise photon rate below 0.1\,Hz. 
\end{abstract}

\maketitle

\section{Introduction}
Photonic interface with stationary quantum systems (quantum memories, matter qubits e.g., atomic ensembles and quantum dots) and quantum state transfer over long distances are fundamental enabling technologies for scalable quantum networks. Quantum networks encompass applications in quantum communication \cite{knaut2024entanglement, yu2020entanglement, van2020long, luo2022postselected, van2022entangling, maring2017photonic}, quantum repeaters \cite{hermans2022qubit, munro2010quantum}, quantum sensing \cite{khabiboulline2019optical}, distributed quantum computation \cite{monroe2014large}, single photon generation \cite{lindner2009proposal}, and deterministic quantum operations between photonic qubits \cite{schwartz2016deterministic}. Quantum memories serving as quantum network nodes have been implemented in various platforms including trapped ions \cite{stephenson2020high,bock2018high, krutyanskiy2019light}, neutral atoms \cite{ikuta2018polarization, van2020long, luo2022postselected}, solid-state color centers such as NV center in diamond \cite{stolk2024metropolitan, stolk2022telecom, dreau2018quantum}, SiV in diamond \cite{bersin2024telecom,schafer2023two,sukachev2017silicon}, rare-earth-doped crystals \cite{lago2021telecom, puigibert2020entanglement} and quantum dots \cite{weber2019two, singh2019quantum}. However, most promising quantum memories operate in the visible or short near-infrared (NIR), making it hard to deploy in low-loss communication fibers. To build compatibility with existing low-loss telecommunication infrastructure, a photon-memory interface is required that connects quantum memories with telecom networks. It can be achieved by quantum frequency conversion (QFC) of visible/near-IR photons into telecom photons, entanglement swapping of quantum memories \cite{lu2019chip}, optomechanical interactions \cite{dong2012optomechanical,hill2012coherent}. However, non-resonant QFC processes have proven to be difficult because of strong classical pump-induced SFWM, Stokes and anti-Stokes noise photons, and laser stability at up- and down-conversion steps  \cite{wang2023quantum,knaut2024entanglement} . Most importantly, they have very low internal conversion efficiencies, and are also limited by losses incurred at fiber coupling leading to poor external conversion efficiencies. This raises the necessity of on-chip resonant solutions that can simultaneously reach higher external efficiencies at low-noise low-pump power and pave the way for scalable integration. Apart from uses in memory-photon interfaces, QFC has applications also in quantum information research, including single photon detectors \cite{liao2017long, ma2012single, pelc2011long}, photon-color erasers \cite{qu2019color,maring2018quantum}.

\begin{figure*} [!ht]
    \centering
    \includegraphics[width=1\textwidth]{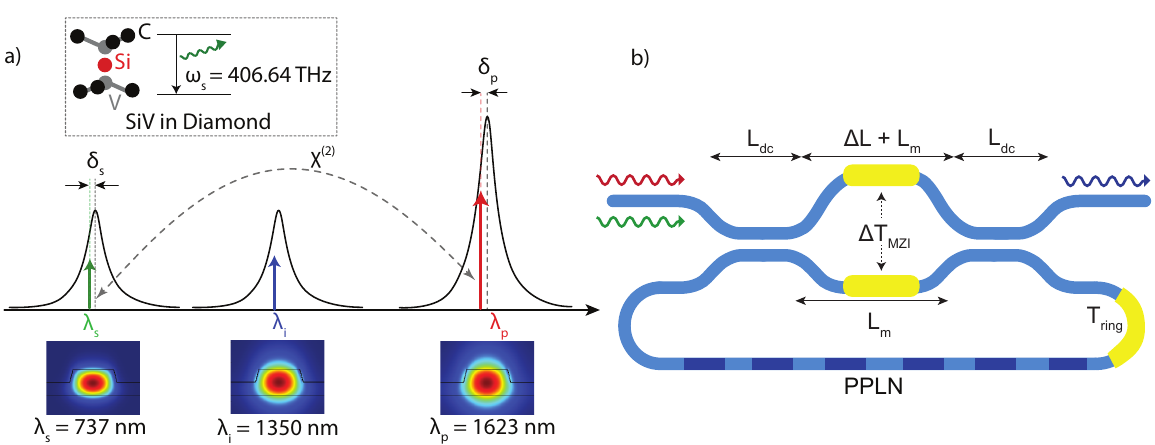}
    \caption{a) Scheme for cavity-enhanced QFC. The classical pump field at frequency $\omega_p$ converts signal photons $\omega_s$ into output idler photons $\omega_i$, dictated by difference-frequency conversion $\omega_i = \omega_s - \omega_p$. Nearest cavity resonances (mode profiles in the bottom) and associated detunings $\delta_s, \delta_i, \delta_p$ are indicated. Phase-matching among the interacting modes is equivalent to the condition on the azimuthal mode numbers $m$ that satisfies $m_s=m_p+m_o+M$, where polling-compensated azimuthal number $M$ is defined as, $M=\frac{L_\mathrm{PPLM}}{\Lambda}$, $\Lambda$ being the polling period. b) Schematic of the PPLN ring resonator coupled with Mach-Zehnder interferometer with asymmetric arms and thermal tuners.}
    \label{fig:Fig1}
\end{figure*}

NIR-to-telecom QFC has been demonstrated based on cascaded $\chi^{(2)}$ process \cite{schafer2023two,saha2023low,fisher2021single}, $\chi^{(2)}$ processes in periodically poled lithium niobate (PPLN) crystals \cite{stolk2022telecom}, four-wave-mixing in chip-integrated $\chi^{(3)}$-microresonators e.g., silicon nitride (SiNx) \cite{singh2019quantum}, aluminium nitride (AlN) \cite{wang2021efficient}. Among these, $\chi^{(2)}$ process is preferable because of the need for lower pump power and the potential for low noise operation. Thin-film lithium niobate (TFLN) is one of the most promising $\chi^{(2)}$ platforms in this regard, offering strong $\chi^{(2)}$ non-linearity (e.g., $d_{33}=25$\,pm/V), fast EO, stable thermo-optic (TO) effect \cite{xu2020high}, and low-loss linear optical characteristics in a broad spectral range. TFLN has proved its potentials in applications ranging frequency comb generation \cite{zhang2019broadband,wang2019monolithic}, efficient light generation ranging from mid-IR to UV \cite{sayem2021efficient}, high purity and efficient single-photon generation \cite{zhao2020high}, electro-optic (EO) modulation \cite{wang2018integrated,xu2022dual}, optical parametric oscillator (OPO) \cite{lu2021ultralow}, squeezing light generation \cite{nehra2022few}, 2nd harmonic conversion \cite{wang2018ultrahigh}.

In this article, we demonstrate the potential of a resonator-coupled Mach-Zehnder interferometer (RMZI) with a PPLN waveguide as an efficient, low-noise, and scalable QFC interface. We chose SiV-in-diamond as our quantum memory which stands out compared to other platforms because of strong coupling to nanophotonic resonators, narrow linewidths and long spin coherence \cite{becker2017coherence} and superior performances in enabling quantum technologies \cite{knall2022efficient, bhaskar2020experimental, nguyen2019integrated}. Our proposed RMZI device can efficiently convert 727\,nm SiV memory transition photon into telecom band for low-loss transmission in optical fiber. Employing temperature tuning in the MZI arms and the ring resonator, we can tune the external coupling rates of the signal, idler, and pump modes and satisfy triply resonant conditions with minimum detuning at the fixed quantum memory wavelength. Using our proposed device, we first show that on-chip internal conversion efficiency can be maximized arbitrarily close to 1 with pump power as low as $\mathrm{\sim 1\,mW}$. By simultaneously fulfilling the over-coupling and the critical coupling conditions for the relevant modes, the external conversion efficiency reaches up to 90\%. Secondly, we derived the four-wave-mixing noise photon rate in the output idler mode and showed the noise count rate to be below 10\textsuperscript{-1}\,Hz, which promises a huge improvement compared to other QFC processes in bulk optics \cite{geus2024low}. Thus, our proposed device offers a fabrication-tolerant solution for a high-efficiency bidirectional QFC process, which can act as an elemental building block for quantum networks. 

\section{Device Concept}
\begin{figure*} [!ht]
    \centering
    \includegraphics[width=1\textwidth]{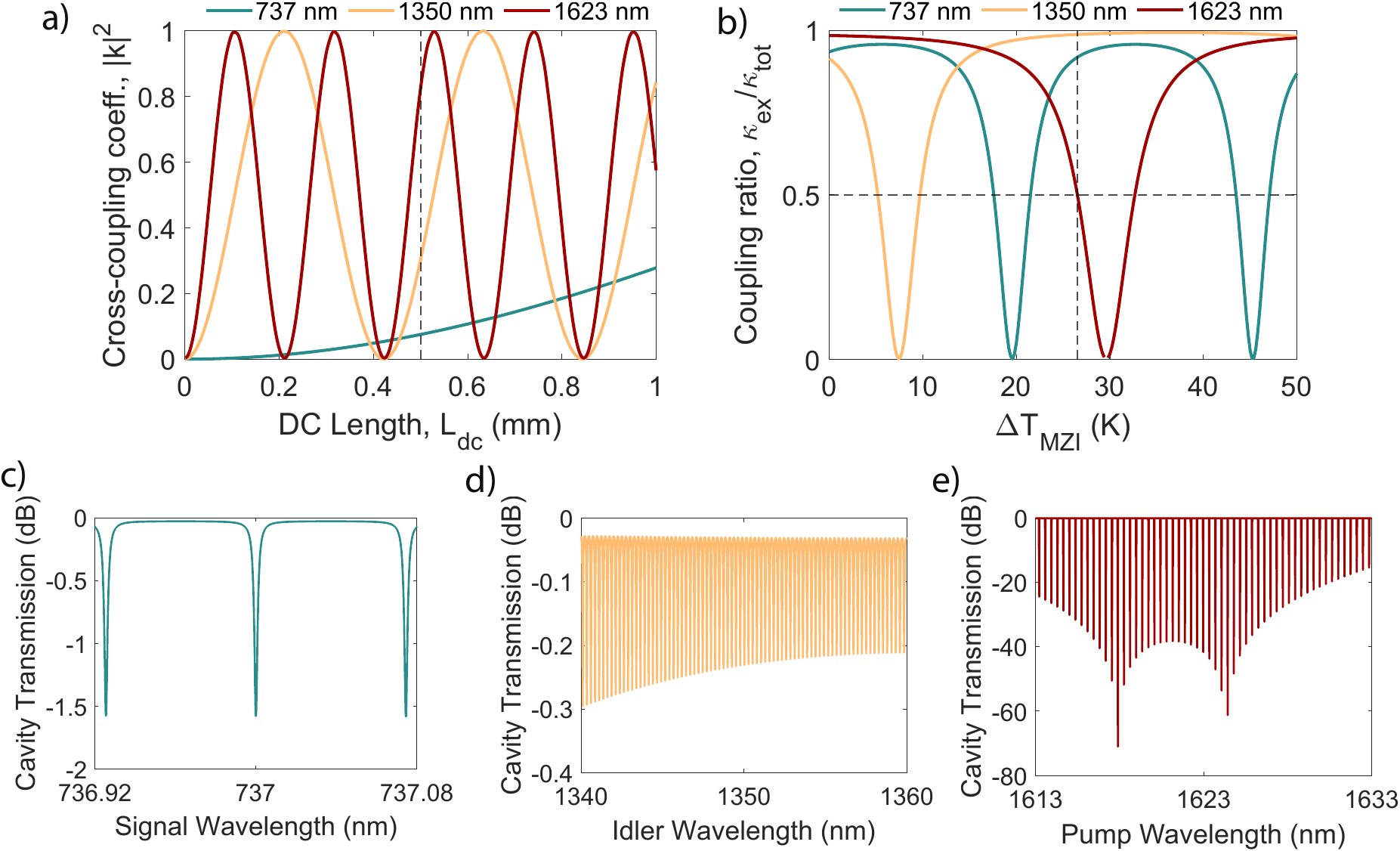}
    \caption{a) Simulated cross-transmission coefficient of the directional coupler, $|k|^2$ for g\textsubscript{dc} = $\mathrm{600\,nm}$ at signal, output and pump wavelengths b) Coupling ratios for the three interacting modes at different MZI temperature with the vertical line marking the operating case. c)-e) show cavity transmissions at the operating couplings as marked in b).}
    \label{fig:Fig2}
\end{figure*}

The primary elements of the device comprised an MZI coupled to a ring resonator with two independent thermal tuners: one tuning section in push-push configuration placed at the MZI arms and another in the ring section. Quasi-phase-matching (QPM) for $\chi^{(2)}$ interaction is attained in the periodically poled lithium niobate (PPLN) section. Details of the cross-sections of the device with thermal heaters and the directional coupler (DC) can be found in our earlier work \cite{kundu2024periodically} and in the supplementary section S1\,\cite{supp}. Fig.\,\ref{fig:Fig1}(a) shows the schematic for the quantum frequency conversion through non-degenerate $\chi^{(2)}$ process, where signal mode $\lambda_s$ is converted to output mode $\lambda_o$ assisted by a strong classical pump field $\lambda_p$. The properties of the signal photon are determined by the SiV transition energy and linewidth, and thus $\lambda_s$ is set by the particular quantum memory technology. In contrast, the operating wavelength of the output and pump modes are flexible as long as they are within the O-band and C-band range for low-loss long-distance propagation i.e. compatible with current telecommunication technology. Fig.\,\ref{fig:Fig2}(a) and (b) depict the cross-coupling transmission for the DC and different coupling ratios achieved by the push-pill thermal tuning of the MZI. The details of thermal tuning mechanism can be found in \cite{kundu2024periodically}. Unless otherwise stated, we choose the top width of the waveguide to be 1500\,nm both for the DC and the ring region throughout the paper. To achieve maximum total or external conversion efficiency at minimum pump power, signal and output modes need to be highly over-coupled, and such condition is achieved at $\Delta T_\mathrm{MZI} \sim 26.5\,K$. To note, slight asymmetry in the upper and lower MZI arm lengths is necessary to facilitate the desired coupling coupling conditions at attainable $\Delta T_\mathrm{MZI}$. We added an asymmetry of $\Delta L \sim 1\,\mu m$ throughout the work. However, asymmetric MZI introduces an envelope response proportional to $\cos^2\left(\frac{\beta \Delta L}{2}\right)$ on top of the cavity spectra, which introduces wavelength-dependent coupling conditions and might complicate frequency-matching at fixed coupling ratio. Thus, it is desirable to keep the asymmetry as small as possible, until appropriate coupling conditions are achieved at the attainable range of $\Delta T_\mathrm{MZI}$. In Fig.\,\ref{fig:Fig2}(c)-(e), cavity transmission spectra for the three interacting modes are shown. 

Satisfying the momentum and energy conservation requirements, we considered the following interaction Hamiltonian for the non-degenerate three-wave mixing process in the resonator,
\begin{equation}
\label{Eq_H}
\mathcal{H_I}=g_0\left(\hat{a}  \hat{b}^{\dagger} \hat{c}  + \hat{a}^{\dagger} \hat{b} \hat{c}^{\dagger}\right)
\end{equation}
where, $\mathrm{\hat{a}} (\mathrm{\hat{a^{\dagger}}})$, $\mathrm{\hat{b}} (\mathrm{\hat{b^{\dagger}}})$ and $\mathrm{\hat{c}} (\mathrm{\hat{c^{\dagger}}})$ are the annihilation (creation) operators for the pump, signal and output mode with frequency $\mathrm{\omega_{a}}$, $\mathrm{\omega_{b}}$ and $\mathrm{\omega_{c}}$ respectively, and $g_0$ is the vacuum coupling rate which depends on the second-order nonlinear coefficient $\chi^{(2)}$ of LN as well as the modal overlap and the geometry of the device. We calculate the $g_0/2\pi$ to be $0.31\,\mathrm{MHz}$ with the PPLN region covering 45\% of the total length of the RMZI resonator. The complete dynamics of the full RMZI PPLN device including intrinsic loss mechanisms and extrinsic couplings through the input and output ports of the upper arm of the MZI is described supplementary. Here, we denote the coupling ratio as $\eta_i = \frac{\kappa_{i,\mathrm{ex}}}{\kappa_{i,\mathrm{tot}}}$, where, $\kappa_{i,\mathrm{ex}}$, $\kappa_{i,\mathrm{0}}$ and $\kappa_{i,\mathrm{tot}} = \kappa_{i,\mathrm{ex}} + \kappa_{i,\mathrm{0}}$ are respectively extrinsic, intrinsic and total loss rate for the pump, signal and idler modes with $i = \mathrm{\{p, s, i\}}$.
\\ 
\section{Quantum Frequency Conversion}
\begin{figure*} [!ht]
    \centering
    \includegraphics[width=1\textwidth]{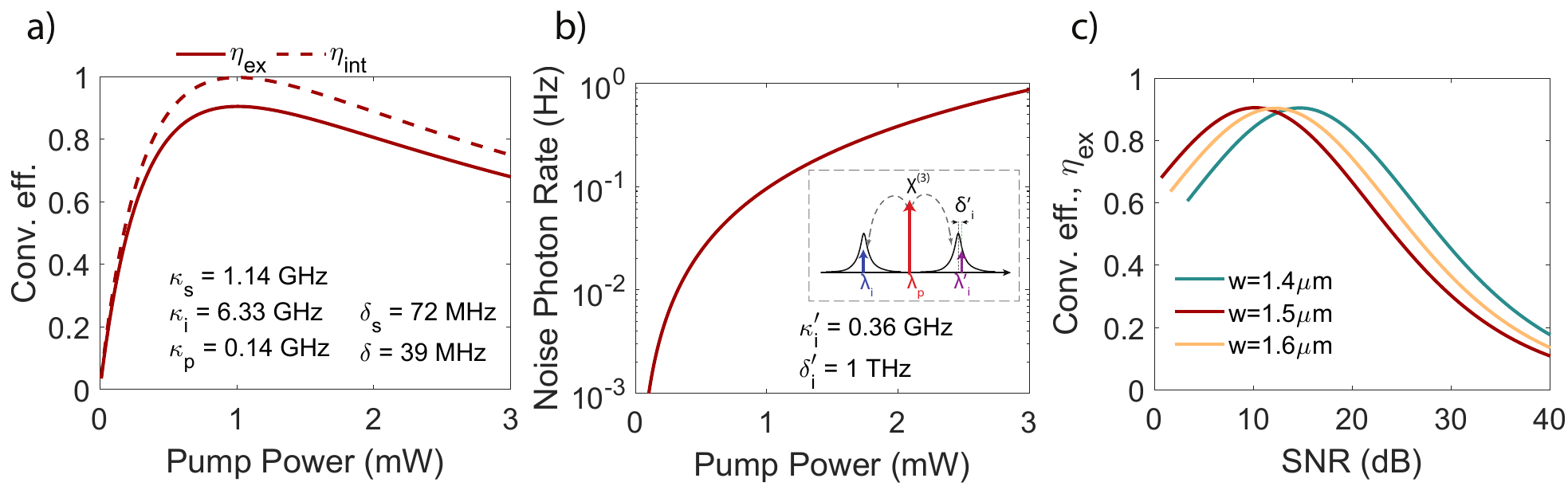}
    \caption{a) Internal and external conversion efficiencies at the coupling ratios set by the $\Delta T_\mathrm{MZI}$ marked in Fig.\,\ref{fig:Fig2}(b) and thermally tuned triple-resonance at $T_\mathrm{ring} = 369.3\,\mathrm{K}$. b) SFWM noise photon rate in the output idler mode associated with QFC process for the conditions same as a). Inset shows the scheme for SFWM process where significant detuing exists for the $\lambda^\prime_i$ mode. c) External conversion efficiencies with respect to acheivable signal-to-noise ratio for different dispersion engineered ring resonators, characterized by the waveguide width $\mathrm{w}$.}
    \label{fig:Fig3}
\end{figure*}

A major challenge of the proposed device is to match frequencies separated over hundreds of terahertz with an accuracy of a cavity linewidth (maximum of around 6\,GHz for the overcoupled idler mode). Furthermore, the requirement of the signal mode resonance to be within 200MHz of the SiV-in-diamond transition linewidth \cite{evans2016narrow} poses another constraint on the triply-resonance condition. To find the triply-resonance frequencies, we sweep the thermal tuner temperature $T_\mathrm{ring}$ from 300\,K upto 400\,K to ensure shift in the base resonance peaks of at least one free-spectral-range (FSR). Setting the acceptable signal mode detuning  to be less 200\,MHz, we search for pump and idler resonance peaks within 20\,nm window around the base wavelengths of 1623\,nm and 1350\,nm respectively. We impose a tight condition on the triply-resonance frequency mismatch parameter $\delta$ to be less than 150\,MHz \cite{evans2016narrow} to cover the narrow linewitdh of critically coupled pump mode. It is to note, we could, in principle, relax $\delta$ upto 6\,GHz set by the overcoupled idler's linewidth and achieve acceptable energy matching. Equipped with frequency-matched resonance mode tuned by $T_\mathrm{ring}$ and desirable coupling ratios set by the $\Delta T_\mathrm{MZI}$, we can directly estimate the external conversion efficiency between the signal and output photon interacting through $\chi^{(2)}$ process, 
\begin{equation}
\label{eta_equ}
    \begin{aligned}
    \eta_\mathrm{ex} &= \frac{\kappa_{s\mathrm{,ex}}}{\kappa_s} \frac{\kappa_{i\mathrm{,ex}}}{\kappa_i} \eta_\mathrm{int} \\
    &=\frac{\kappa_{s\mathrm{,ex}}}{\kappa_s} \frac{\kappa_{i\mathrm{,ex}}}{\kappa_i} \frac{4C}{\left|\left(1 + i\frac{2\delta_s}{\kappa_s}\right) \left(1 + i\frac{2(\delta_s - \delta_p - \delta)}{\kappa_i}\right) + C\right|^2}
    \end{aligned}
\end{equation}

Where, $C = \frac{4g^2|\alpha|^2}{\kappa_s \kappa_i}$ is the system cooperativity and $\delta = \omega_b - \left(\omega_a + \omega_c\right)$ is the frequency mismatch for the triply-resonance condition in the resonator. $\delta_s$ and $\delta_p$ are the detuings respectively for signal and pump modes. According to Eq.\,\ref{eta_equ}, the maximum internal conversion
efficiency will be achieved when $C = 1$ without any detuning, which corresponds to a drive pump power, 
\begin{equation}
\label{Pmax_equ}
    \begin{aligned}
    P_\mathrm{max} = \left[\frac{\kappa_{p\mathrm{,0}}\kappa_{s\mathrm{,0}}\kappa_{i\mathrm{,0}}}{16 g_0^2 \eta_p (1-\eta_p)(1-\eta_s)(1-\eta_i)}\right] \hbar \omega_p
    \end{aligned}
\end{equation}

From Eq. \ref{eta_equ}, it is evident that, even when $\eta_\mathrm{int}$ reaches its maximum value of 1- corresponding to $C = 1$ and zero detuning, the external conversion efficiency is still limited by the coupling ratio of the signal and idler mode. To maximize $\eta_\mathrm{ex}$, theoretically, these coupling ratios need to be unity. However, such strong overcoupling of the signal and idler modes constrains on the pump power required for unity cooperativity. From Eq.\,\ref{Pmax_equ}, it can be observed that the pump power for unity cooperativity, $P_\mathrm{p}$ depends on the coupling conditions of all three modes, but in distinct ways. For pump mode, it needs to be critically coupled to maximize the product $\eta_p(1-\eta_p)$. Conversely, $P_\mathrm{p}$ has an inverse relationship with the signal and idler coupling ratios, $\eta_i$ and $\eta_s$. Consequently, strong overcoupling of signal and idler modes increases the required pump power. This results in a trade-off between maximizing $\eta_\mathrm{ex}$ and minimizing drive power requirement. As mentioned earlier, our proposed RMZI with a thermal tuning scheme can achieve the sweet spot where the pump is critically coupled, and signal and idler are overcoupled (Fig\,\ref{fig:Fig2}(b)). Consequently, the RMZI can achieve a maximum value of $\eta_\mathrm{ex}$ as high as 0.9 at around 1 mW power, which is quite high compared to the reported values in the literature at a comparatively low power \cite{wang2023quantum,knaut2024entanglement}. Both $\eta_\mathrm{int}$ and $\eta_\mathrm{ex}$ are depicted in Fig.\,\ref{fig:Fig3}(a) for different pump powers for unity cooperativity. With higher pump power, the conversion efficiencies decrease. This can be attributed to the redistribution of increased power towards the side-band resonant modes, diverting the energy away from the desired signal/idler modes, thereby reducing the efficiency. Thus, the proposed system can achieve a very high external frequency conversion efficiency at almost sub mW power. 

\section{Pump-induced Noise Calculation}
As discussed in the previous section, the efficiency decreases with increasing pump power owing to the redistribution of energy towards the side band resonant modes. This energy shift can result in the generation of noise photons at undesired modes during the frequency conversion process \cite{wang2023quantum,knaut2024entanglement}. Furthermore, at very high pump power, additional non-linear processes, such as four-wave mixing due to $\chi^{(3)}$ non-linearity can also become significant. Therefore, competition emerges between the $\chi^{(2)}$ and $\chi^{(3)}$ processes, where the photons generated by the $\chi^{(3)}$ process contribute to noise photons, thereby reducing the overall efficiency. In this section, we'll analyze the robustness of our system against such pump-induced noise generation. The noise photon generation rate $R_\mathrm{FWM}$ resulting from FWM is given by (details in the supplementary \cite{supp}),
\begin{equation}
% \label{Req}
    \begin{split}
    R_\mathrm{FWM} &=   64g_{\chi^{(3)}}^2\left[\frac{P_p}{\hbar\omega_p}\right]^2\frac{\kappa_{p{\mathrm{,ex}}}^2}{\kappa_{p}^4} \frac{(\kappa_{i}+\kappa_i^\prime)}{4\delta_i^{\prime 2} + (\kappa_{i}+\kappa_i^\prime)^2}.
    \end{split}
\end{equation}
The detailed parameters are provided in the supplementary information. As the primary objective of the QFC process is to minimize detuning for the desired individual modes, the noise photons generated due to the undesired process $\chi^{(3)}$ are largely detuned. In our proposed scheme, the phase-matched (azimuthal mode number) resonant mode resulting from the FWM process exhibits a detuning of $\delta_i^\prime = 1$\,THz. More specifically, through the pump-induced FWM process, two pump photons are generating one noise photon in the output resonant mode in O-band, and one lower frequency photon in a phase-matched mode supported by the resonator. Moreover, under this condition, lower frequency idler mode is slightly overcoupled with $\kappa_i^\prime = 0.36$ GHz. The derived noise photon generation rate with varying pump power is shown in Fig.\,\ref{fig:Fig3}(b). With increasing power, $R_\mathrm{FWM}$ increases as we expect, but the noise photon rate is very low ($<< 1$ Hz) even at high power. In fact, at around 1 mW pump power, the noise photon rate is below $0.1$Hz, a point where the system achieves the maximum conversion efficiency. Since the noise photon generation rate is critically dependent on the value of detuning $\delta_i^\prime$ of the phase-matched idler mode, we opt for engineering the dispersion profile of the resonator. Keeping the coupling ratios unchanged as depicted in Fig.\,\ref{fig:Fig2}(b), we optimize the frequency detunings for ring widths of 1.4\,$\mu$m and 1.6\,$\mu$m, resulting in ring-tuner temperature $T_\mathrm{ring}$ of 364.1\,K and 312.5\,K at pump resonance 1631.4\,nm and 1630.1\,nm respectively. In Fig.\,\ref{fig:Fig3}(c), we plot the external conversion efficiencies of the QFC process for different signal to $\chi^{(3)}$ noise ratios, ($\mathrm{SNR\,(dB)} = 10\,\mathrm{log}R_\mathrm{FWM}$,  achievable by varying the pump power. Ring resonator with a width of 1.4\,$\mu$m provides higher SNR at the maximum $\eta_\mathrm{ex}$ compared to other ring widths, which proves the provision of dispersion engineering to further improve the noise photon rate.

\section{Conclusion}
In conclusion, we show that using an asymmetric ring-Mach Zehnder interferometer coupled with a PPLN waveguide, it is possible to convert photons at the quantum level from quantum memories to low-loss telecom channel with external conversion efficiency as high as 90\% at mW pump power with noise photon rate below 0.1 Hz. Our proposed design provides a significant step forward in realizing true quantum interconnects connecting quantum memories to long-distance telecommunication channels. 

\section{Author Contribution}
A.S. conceived the original idea. M.K, B.S. performed the theoretical and numerical simulations with supervision from A.S. M.E. and A.S. reviewed the technical details.  

\bibliography{Reference}

% \onecolumngrid
% \input{Supplementary}

\end{document}